# Attractors, Equilibrium, Profit, Loss

by Leonid A. Shapiro

**Abstract.** Various formulations of counterfactual general equilibrium in economies — systems of actors manipulating economic goods — are logically and mathematically analyzed. Evenly-rotating economies are systems whose evolution is stable, steady, and form-preserving. Several of their qualities are qualities of « homeorhesis » but they are not in « homeostasis »: because their instantaneous state is not steady, or in other words, they are not « stationary ». What is possibly stationary in whole economies is change over time of behavior of their parts in a way that causes that net of relations of these parts which results to be preserved over time. Necessary existence of actual disequilibrium and impossibility of actual or ultimate equilibrium is demonstrated and different types of equilibrium and disequilibrium are discussed.

## §1
### Introduction

« Dynamics » is analysis of states of systems of bodies and forces and it consists of « statics » — analysis of states of such systems whose states are not changing over time — and « kinematics » — analysis of states of such systems whose states are changing over time (Tait & Thomson 1867). For instance: those systems whose states are changing over time are « evolving » — changing non-periodically and regularly — non-singularly — while they transition or their state converges to a future state that satisfies one or several constraints which are not already satisfied — or developing or growing or reproducing (Spencer 1864; Lotka 1925; Waddington 1957).

« Equilibrium » is steadiness of each state whose null change — once it already results — is a possible change such that all changes that possibly or actually take place in space and time are changes that cause states of dynamical systems which are distinguishable every one from every other one by measurement of their distances or divergences from that steady state. Thus exist distinguishable equilibria, and statics is analysis of equilibria of dynamical systems, but what is an equilibrium of those dynamical systems which are called « economies » ?

These systems are sets which consist of one or several distinguishable actors each of whom is individually manipulating marginal quantities of economic goods — concrete parts of his or her mind's future-past-and-present exterior which he or she possibly gains or loses at one or other moment of time as wholes which he or she prefers to experience in a concrete order that itself possibly changes over time — from his or her mind's perspective in space-time.

All economies are thus special dissipative systems — they are « open » systems — which is to say that they are systems which are fundamentally coupled with parts of their exterior and these parts are altogether called future-past-and-present « environment » of these systems.

Economies in state of equilibrium by way of hypothesis are not themselves « stationary ». Their instantaneous state is not steady. Economies in state of equilibrium by way of hypothesis are not themselves in « homeostasis », which is a mode of existence that is not changing.

If their state is steady while state of their environment itself does not change, then contradiction is absent in many dynamical systems but such instantaneous or momentary homeostasis cannot exist in economies: they are themselves aggregates of actors who manipulate — consume and produce — economic goods ceaselessly over time. Each moment these actors have some wants which are not already satisfied — their wants are not limited in number while they live and so they always remain in motion: they behave in that way which they believe — expect — shall cause them to experience events in that one temporal order which they prefer, namely, which shall satisfy their desires. Without actors — which are necessarily living actors — economies cannot themselves possibly or actually exist; but actors remain in motion while state of their environment changes and necessarily too they remain in motion when it ceases changing. Economies are living systems: because they consist of actors who themselves are living systems. If state of economies is steady while state of their environment changes, then they are not living systems, which is contrary to hypothesis that they are economies: their exterior impinges on their form — their organization — and disturbs it or destroys it — and so these systems themselves — except if and only if their internal state already compensates for future state of their exterior prior to it replacing present state of their exterior and in order to do this they must presently anticipate their possible or actual future environment (Spencer 1864; Rosen 1985; 1991).

That which is possibly steady or not steady in whole economies is change over time of behavior of actors who are their parts. If such change is steady and stable and it preserves form of economies — which is momentary set of relations of their parts over time — while state of their environment varies, then « even-rotation » of these economies results (Mises 1949). Indeed environment of evenly-rotating economies changes because behavior of actors which is anticipated to satisfy their wants or actually does so necessarily changes it; action is behavior and like no effect exists without a cause so no cause exists without an effect; but truly even-rotation of economies is possible while state of their environment is not steady. Whichever observer believes or supposes that environment of evenly-rotating economies is not changing thus vainly postulates existence of superfluous causes of such even-rotation: he or she postulates existence of those causes which are neither sufficient nor necessary for such even-rotation to take place but such causes cannot exist in nature and postulating them cannot make true causes of phenomena — series of those events which observers feel — known to observers. Let us therefore not assume steadiness of environments of economies.

Equilibrium is steadiness of each state whose steady change is a possible change such that each change that possibly or actually takes place in space and time is a change which is measured relative to that special state and so exist distinguishable equilibria. For instance: observe that homeostasis — steadiness of state of systems themselves — is equilibrium of many systems in equilibrium but true homeostasis of economies is not possible.

All economies are ensembles of fundamentally coupled or linked quantitative variables and variable quantities, namely, they are complex systems, which is to say, in other words, that relations of these quantitative variables and variable quantities cannot themselves be reduced — be coherently made identical while preserving behavior of systems which they generate — to sets of relations which are fewer in number than they or which lack this coupling (Rosen 1968).

All modes of existence of systems are distinguished from other such states by a process of measurement of distance or divergence) of such from another state, a basis of measurement, and so change and only change of state is perceptible itself from perspective of systems (Boskovich 1755).

If states which are bases themselves vary over space and time by hypothesis, then divergence from them cannot by process of identifying names or things which are identical to identical names or things itself distinguish states over space and time from perspective of those systems who observe that such-and-such states diverge so-and-so from these bases; but that is contrary to hypothesis that states which are bases are truly bases and states are distinguished by measurement of divergence from them, which is self-contraction, impossible, and so bases of measurement are steady over space and time by necessity (Hutton 1794a,b).

Equilibrium cannot be steadiness itself there where this is a basis of measurement. That steadiness which causes it really is a stable steadiness: because there where infinitesimal change of causes itself causes greater than infinitesimal change of their effects in space of all possible such causes and their effects distances or names of several effects cannot be determined or inferred coherently according to distances of their causes in that space (Thom 1983).

Equilibrium of those simple systems which are not themselves alive exists there and « only there where state of their entire motion is already a steady state », or in other words, « qui tamen tantum cum totus motus sese jam ad statum permanentem composuerit » (Euler 1742).

This is not itself possible in living systems which consciously or purposefully behave — in these systems which move thus how they anticipate causes them to remain alive or causes satisfaction of their wants — because ceaseless motion of actors in order to satisfy their wants which are not already satisfied — and such wants are infinite at every moment — and change of their environment caused by their movement or other forces and their responses to such change causes all quantities which are parts of motion or state of actors to change over time in economies: constant quantities observed in simple systems are absent in economies, which causes true uniqueness over time of each actual state of economies (Mises 1949; 1957). Inference reveals that which is true: and what is that? A steady change of behavior or motion of actors in economies — steady change of change of state of those economies — is that one steadiness that is truly possible in economies.

This steadiness of change of behavior of systems is possibly stable or it is not <u>so</u>; it possibly preserves their form or <u>this</u> it does not do, or it possibly preserves their behavior, too, or <u>this</u> it does not do; it possibly causes their behavior to be stable or <u>this</u> it does not do; it possibly causes their homeostasis or <u>this</u> it does not do. None of these possible qualities of steady change of behavior of systems are themselves necessary qualities of such change. Equilibrium of economies or other living complex systems is steady form-preserving stable change of their behavior, which is to say, in other words, that it is their even-rotation, and why? They are living systems; change of their behavior preserves their form. If this it does not do, then they cease to exist, which is contrary to hypothesis that they exist in equilibrium, and there where divergence of other change of their behavior from this change of their behavior is measured they are bases of measurement, which implies that equilibria are stable.

Thus several of even-rotation qualities are qualities too of « homeorhesis », which is steady form-preserving stable change of state of systems (Waddington 1957), and where these special equilibria actually or possibly exist they are there caused by « life », which is one-or-several-steps-ahead correspondence of state of interior of systems and state of their exterior — correspondence of state of systems themselves and their environment — which is possible itself if and only if these systems anticipate possible future state of their exterior and respond presently or some other time which is prior to it taking place (Spencer 1864; Rosen 1985; 1991).

Observe this principle: life is possible without even-rotation or homeorhesis but even-rotation and homeorhesis are not possible without life. In fact homeorhesis exists primarily during <u>pre-determined</u> <u>non-conscious</u> development or growth of living organisms (Waddington 1940; 1957).

How can form of economies be preserved during action or interaction of their parts? That which preserves their form must also cause at once stability and steadiness, too, of change of action or interaction of these parts: one and only one cause of all this is possible. Equilibrium of economies is steadiness of change of behavior which preserves their form because actors which are their parts already somehow anticipate all possible and actual future change of their environment, including that which they cause, and their own future wants (Wicksteed 1910; Mises 1949).

Sets of actors who voluntarily associate — because each person satisfies wants not already satisfied and so « gains » there where each person of every two actors voluntarily gives in exchange quantities of goods which he or she feels are superfluous or less useful in process of satisfying his or her wants than quantities of goods which he or she receives in exchange — are « societies » and ensembles of locations in space-time where these co-operations take place are « markets » (North 1691; Bonnot de Condillac 1776; Destutt de Tracy 1817; Longfield 1834; Gossen 1854). Thus societies and markets whose behavior is steady are necessarily not themselves in homeostasis: once all possible production or exchange of goods by actors whose preferences are opposite in a partition their preferences is exhausted societies and markets themselves necessarily cease to exist (Gossen 1854; Wickstead 1910; Mises 1912; 1949): but that is contrary to hypothesis that behavior of societies and markets is in state of homeostasis.

Those dynamical systems which cannot possibly exist cannot possibly have behavior and so evenly-rotating economies consist of actors whose preferences and properties are such that they can mutually gain by voluntarily exchanging their some of their property: consumption, exchange, and production all take place during such even-rotation and behavior of actors doesn't change over time but actors themselves can and do change state over time. What is truly absent while actors behave so is « uncertainty »: all possible future events which actually take place are somehow perceived by actors arbitrarily earlier than moments when these events take place from perspective of actors by way of some process of anticipation and those events which actually do not take place are not anticipated by actors to do so. (Mises 1949)

An event is « anticipated » if it is judged to happen in the future by some person. That event is « expected » if personal experience or sensation of it by him or her who anticipates it is itself anticipated too: he or she anticipates that he or she later somehow observes this event. A person buys that or this good: he or she expects that it satisfies his or her future wants or he or she expects that it can be later exchanged for something that would satisfy his or her future wants.

L.A. Shapiro	March 21, 2013	page 6

Profit and loss are indices which are derived from buying or selling operations which take place amidst consumption or production operations by way of calculation of distance in phase-space of actual state of an economy from its counterfactual equilibrium.

These indices measure divergence of future state of an economy that is expected by actors who are its parts from actual future state of that economy. Profit and loss are perceived only in kinematics of change of instantaneous behavior of actors in an economy that consists of them but these indices are necessarily measured relative to counterfactual even-rotation of that economy while it is perceived momentarily by way of hypothesis in statics of change of its instantaneous behavior (Mises 1949; 1952; Hulsmann 2000).

How are these indices calculated? Positive difference of money-revenue and money cost — « profit » — occurs by way of some actors who are called « entrepreneurs » purchasing goods at known and thus certain prices and selling later at prices which are not certain from perspective of those actors when they purchased goods because true change over time of preferences of consumers or rarities of factors of production cannot be certainly known from perspective of actors themselves within economies. Negative difference of money-revenue and money-cost — « loss » — is caused by sale of goods at money-prices that were not certain during their production or purchase from perspective of actors too. (Cantillon 1755; Mises 1949; 1952)

A common medium of exchange called money — something that is easily exchanged and whose quantities are ranked in preferences of <u>all</u> actors — is held by every person: in order to satisfy want of ability to trade property where parties exchanging it do not have preferences are not opposite but parties can give money to yet other parties in exchange for yet other economic goods they do actually want — because money is more frequently saleable than all other commodities — and in order to satisfy want of ability to certainly later buy economic goods whose purchase is not presently anticipated but possible — by way of holding money to give in exchange for them (Gossen 1854; Menger 1871; Mises 1912; 1949; Hutt 1939; 1952).

Quantities of money thus directly satisfy wants and so they are first-order goods which are ultimately ranked amongst other first-order goods in preferences (Mises 1912). Without money-prices value of higher-order goods — which cannot directly satisfy wants and so are not ranked in preferences — cannot be compared with value of lower-order goods: money-prices make possible calculation of quantity over time of factors of production which is required in order to satisfy greatest number of highest-ranked wants and so money has this use, too, once it already exists, but this use derives from value of money and is not source of its value (Mises 1949).

One may argue: « general equilibrium cannot exist, because we cannot conceive of a world where every person has perfect foresight and where no data ever change, and it is self-contradictory: one holds cash balances because of uncertainty of the future and therefore demand for money is zero in a general equilibrium world of perfect certainty, hence a money economy could not be in general equilibrium. » (Rothbard 1979)

Actually: some quantity of money greater than zero would necessarily be demanded because not every source of desire to hold money is annihilated by way of hypothetical absence of uncertainty, or in other words, not every source of desire to hold money is annihilated by way of hypothetical ability of actors to anticipate every future change of their present environment (Mises 1949).

If actors neither hold money nor periodically exchange quantities of it for some goods, then which one in every pair of first-order goods is preferred by consumers and which one in every pair of higher-order goods contributes more to satisfaction of preferences of consumers — information that is itself dispersed among consumers and producers — is not indexed by money-prices and thus it is not made available to all producers who therefore cannot decide except arbitrarily or by way of guessing which lower-order goods to manufacture, when, in what quantity, and at what cost. Owners of higher-order goods cannot then decide except by way of guessing, too, which one in every pair of possible but not co-possible employments of their property to realize and which one to forego because they lack an index of extent that these alternative actions contribute to satisfaction of preferences of other persons and so do not know what they must expect to be given in exchange for use of their property and therefore ultimately they do not know which possible use of their property they must forego in order to satisfy their own preferences to greatest possible extent. If hypothetically each person already perceives all possible events which actually take place somewhere in space-time, e.g., preferences of consumers, knows which events do not do so, and knows least-cost procedures of satisfying wants, or somehow truly anticipates this without information, then neither such indexing nor knowledge which it conveys is necessary in economies: money thus loses one source of its value. (Gossen 1854; Hayek 1945)

Thus demand for money in evenly-rotating economies is less than demand for money in non-evenly-rotating economies which hypothetically consist of actors identical to actors comprising those evenly-rotating economies; their even-rotation annihilates one source of desire to hold money; but not every source of desire to hold money is so annihilated (Mises 1949; Rothbard 1979).

What phenomena necessarily cease to exist where uncertainty is hypothetically absent? Profit and loss are totally annihilated by even-rotation of economies by way of total absence of uncertainty in such economies: preferences of consumers or rarities of higher-order goods productively consumed in order to make products do not change over time during even-rotation and so consumption of product by its consumers and its production occurs at various frequencies and these coincide during even-rotation of economies where consumption and production take place. If so, then money-revenue neither exceed money-cost over time, which results in observable absence of profit, nor money-costs exceed money-revenue over time, which results too in absence of loss. (Mises 1949)

Entropy of change of behavior of evenly-rotating economies — logarithm of number of distinguishable ways that this state of behavior can possibly result, namely, logarithm of number of symmetries which generate it — neither increases nor decreases. How does this manifest in economies themselves?

Entropy of economies themselves <u>can</u> possibly increase or decrease during their even-rotation. Why? Each system more frequently evolves from lesser entropy state — which is less probable — to greater entropy state — which is more probable — than it evolves from greater entropy state to lesser entropy state. In every system cause equals effects — « energy » is conserved (Leibniz 1695; Mayer 1842) — or in other words, « laws of nature », which « determine » its configuration, are uniform over space and time from every perspective (Hutton 1794a,b). Therefore each system is found in some one state many more times than it is found in some other state — its each state or mode of its existence occurs infinitely many times during infinite space-time — but then each state has some time-less frequency — probability — or number of times that it occurs in all space-time relative to number of times which it or a state different from it so occurs (Lotka 1925) and this because some infinities are different from other infinities by way of absence of one-to-one correspondence their parts (Leibniz 1676).

Each actor plans and anticipates his or her future consumption and production activities: in those economies which are evenly-rotating all such plans of all individual persons are possible and co-possible, internally consistent and mutually consistent, namely, coherent (Hayek 1937).

This is a « test » of even rotation of economies, which is to say, or in other words, that those economies which are in state of « disequilibrium » consist of some entrepreneurs; they also consist of several actors — some of whom are possibly entrepreneurs themselves — whose planned consumption and planned production taken together is ultimately not coherent. So too absence of entrepreneurs in evenly-rotating economies by way of hypothetical absence of a future whose state is « uncertain » from perspective of actors is one other test of even-rotation — « equilibrium » — of all economies which are hypothetically observed and mentally analyzed.

Other counterfactual equilibria of economies can be constructed. For instance, meta-equilibrium is steadiness of non-market social institutions and social organization which condition all change of behavior of economies and so contribute to determining counterfactual or real differences in profit and loss in various economies which already have profit and loss but neither profit nor loss themselves are determined relative to meta-equilibria. This is so because meta-equilibria are themselves extra-market phenomena by hypothesis. Several writers had suggested that statics is analysis of market-equilibria and meta-equilibria too.

For instance, society which is most changeless must necessarily remain in motion while it exists in living state like societies which are not changeless: « among individuals composing it, motion, action, life, adjustment, and effort toward equilibrium exist, and equilibrium must constantly be disturbed by environmental changes and fluctuations in personal qualities of actors. Fundamental contrast is that between stationary and progressive social organizations. In a workable static society, there must be individual activity, forces, movements, children are born, actors grow old, they succeed or fail, advance or decline in individual fortune. Essential mark of a static society is not crystallized human units but fixity of its general form and its social institutions. » (Fetter 1910)

In this essay we shall analyze market-equilibria and not meta-equilibria.

A question can be asked here: can we prove necessary existence of disequilibrium in economies and impossibility of them existing in actual or ultimate state of equilibrium?

## §2
## Notation

For conciseness, because composition appears here more frequently than multiplication, left-to-right-combination juxtaposition is composition but multiplication is explicitly written: $E \circ \iota F = E(\iota)F = E\iota \cdot F$. Identity functor $J = \{\iota \mid F\iota = F = \iota F\} = JJ = J'$; it is that functor that is its own inverse $J' = \{\iota \mid J\iota = J = \iota J\}$. Thus $J \neq F \neq F' \neq J$. (Menger 1944; 1952; 1953)

An ensemble is a set which consists of more than one element.

For instance: a combination $(J \mapsto F)(R \in X) = (J(R \in X) \mapsto (F(R \in X) = ((FR = P) \in (FX \subseteq Y))) = (R \xmapsto{F} P) = (X \xrightarrow{F} Y)$ is part of a category $I \xrightarrow{E} X \xrightarrow{F} Y \xleftarrow{\Gamma} I$ which « commutes » if and only if $EFI = FX = \Gamma I$ (MacLane 1971). Let $(((AB)(EF))((GH)(XY)))Z = AB.EF : GH.XY \therefore Z$ (Peano 1889). There where this is necessary in context $A / B$ is really set $A$ excluding set $B$ instead of quantity $A$ divided by quantity $B$. Let $\text{dom } U = \text{dom } V = \text{dom } W \mid U = V \neq W \mid \text{cod } U = \text{cod } V = \text{cod } W$.

All those natural things which are not themselves marked by one name $X = XY \cup Xy \cup (\emptyset = XY \cap Xy)$ are each one really marked by a name $x = \text{non-}X$ (DeMorgan 1860; Jevons 1864).

If $P \in \{\eta, \ldots, \mu\} = L$, then $(\zeta J \xi_P) = \bigcap_P (\zeta J \xi_P) = \zeta J \xi_\eta \cap \ldots \cap \zeta J \xi_\mu$.

Let $\text{op} = 0 - J$. If $i = \sqrt{\text{op } 1}$ and $2 \cdot \pi \cdot i = j$. If $(\tau, \sigma) \in \mathbb{R} \times (\mathbb{R}/\{0\})$ and $(\tau + i \cdot \sigma, \tau - i \cdot \sigma) = (\kappa, \overline{\kappa}) \in \{\iota \in \mathbb{C}/\tau\} \times \{\overline{\iota}\}$, then « hyper-function » $Ц\, \tau = \lim_{\sigma \to 0+} \{Ц_+ \kappa - Ц_- \overline{\kappa}\} = \text{hyp}(Ц_+ \kappa, Ц_- \kappa) = \text{hyp}(Ц_+, Ц_-)\kappa$ (Kothe 1952; Sato 1958; 1959; 1960).

Let $\mathbf{X} \subseteq \mathbf{Y} \mid \mathbf{R} \subseteq \mathbf{P}$. For instance: this is true in events where $\mathbf{R} < \mathbf{P} < \mathbf{X} = \mathbf{Y}$.

If also $\mathbf{X} = \sum_{V=1}^{T} X_V \cdot \mathbf{\varepsilon}^V \Big|_{X_V \leq Y_V} \in \underset{V=1}{\overset{T}{\times}} \mathbb{R} \Big|_{T \leq I}$ and $\mathbf{Y} = \sum_{V=1}^{I} Y_V \cdot \mathbf{\varepsilon}^V \in \underset{V=1}{\overset{I}{\times}} \mathbb{R}$,

then: $\vartheta_{\mathbf{X}-\mathbf{R}\mid\mathbf{P}-\mathbf{Y},1} = \begin{cases} 1 \text{ if } \mathbf{R} < \mathbf{X} . \cap : \mathbf{P} < \mathbf{Y} . \\ 0 \text{ if } \mathbf{R} < \mathbf{X} . \cup : \mathbf{P} < \mathbf{Y} . \end{cases} = \begin{cases} 1 \text{ if } \mathbf{X}-\mathbf{R} > \mathbf{0} . \cap : \mathbf{Y}-\mathbf{P} > \mathbf{0} . \\ 0 \text{ if } \mathbf{X}-\mathbf{R} < \mathbf{0} . \cup : \mathbf{Y}-\mathbf{P} < \mathbf{0} . \end{cases}$

$= \begin{cases} 1 \text{ if } \mathbf{X}-\mathbf{R} > \mathbf{0} \\ 0 \text{ if } \mathbf{X}-\mathbf{R} < \mathbf{0} \end{cases} \Big| \mathbf{R} \in \underset{V=1}{\overset{T}{\times}} \mathbb{R} - \begin{cases} 1 \text{ if } \mathbf{P}-\mathbf{Y} > \mathbf{0} \\ 0 \text{ if } \mathbf{P}-\mathbf{Y} < \mathbf{0} \end{cases} \Big| \mathbf{P} \in \underset{V=1}{\overset{I}{\times}} \mathbb{R}$

$= \prod_{V=1}^{T} \begin{cases} 1 \text{ if } X_V - R_V > 0 \\ 0 \text{ if } X_V - R_V < 0 \end{cases} \Big| (X_V, R_V) \in \mathbb{R} \times \mathbb{R}$

$- \prod_{V=1}^{I} \begin{cases} 1 \text{ if } P_V - Y_V > 0 \\ 0 \text{ if } P_V - Y_V < 0 \end{cases} \Big| (P_V, Y_V) \in \mathbb{R} \times \mathbb{R}$

$= \prod_{V=1}^{T} \text{hyp} ( \text{op log op} [ X_V - R_V ] / j , \text{op log op} [ X_V - R_V ] / j )$

$- \prod_{V=1}^{I} \text{hyp} ( \text{op log op} [ P_V - Y_V ] / j , \text{op log op} [ P_V - Y_V ] / j )$

$= \prod_{V=1}^{T} \vartheta_{X_V - R_V, 1} - \prod_{V=1}^{I} \vartheta_{P_V - Y_V, 1} = \vartheta_{\mathbf{X}-\mathbf{R}, 1} - \vartheta_{\mathbf{P}-\mathbf{Y}, 1}$

For instance: $\mathbf{X}$ is a set of inputs and $(\rho \mathbf{X}) \varphi \mathbf{X} = \mathbf{Y}$ is a set of all those inputs plus all outputs which are labeled by invertible map $\rho$ after they are associated with them by process $\varphi$, namely, configuration $\mathbf{X}$ corresponds to point-location $\mathbf{Y}$ in a phase-space having dimension $I$. If $\mathbf{R} < \mathbf{P} < \mathbf{X} = \mathbf{Y}$, then $\vartheta_{\mathbf{X}-\mathbf{R}\mid\mathbf{P}-\mathbf{Y},1}$

$$= \prod_{V=1}^{T=I} \text{hyp} \left( \log \frac{R_V - X_V}{Y_V - P_V}, \log \frac{R_V - X_V}{Y_V - P_V} \right).$$

Let $\mathbf{R} < \mathbf{X} = \mathbf{Y} < \mathbf{P}$. If so, then $\vartheta_{\mathbf{X}-\mathbf{R}\mid\mathbf{Y}-\mathbf{P},1}$

$$= \prod_{V=1}^{T=I} \text{hyp} \left( \log \frac{R_V - X_V}{P_V - Y_V}, \log \frac{R_V - X_V}{P_V - Y_V} \right) \text{ (Sato 1959)}.$$

We thus can discuss discrete change and discrete events analytically and so strictly continuously.

## §3
## Analysis

All behavior is taking or leaving, doing or not doing this or that. Processes are completed when actions which are sufficient for their completion are done — when $\vartheta_{X-R\,|\,P-Y,\,1} = 1$.

Topological ball $\wp(J\,;\,\mathbf{C},P,\Omega) = \{\,J\,|\,\Omega(\Omega(J,\mathbf{C}),P) \leq 0\,\}$ has « center » $\mathbf{C} \in \underset{R=1}{\overset{B}{\times}} \mathbb{R}$ and « radius » $P \in \mathbb{R}$ there where $\mathrm{dom}\,J = \mathrm{dom}\,\underset{R=1}{\overset{B}{\sum}} J_R \cdot \varepsilon^R = \underset{R=1}{\overset{B}{\times}} \mathbb{R}$ ; thus its « boundary » $\partial\wp(J\,;\,\mathbf{C},P,\Omega(J_M,J_N)) = \{\,J\,|\,\Omega(\Omega(J,\mathbf{C}),P) = 0\,\}$ is a sphere.

In speech or writing, by convention, « a » is « some one » and « each » is « every one », which is to say, for instance, that each $E$ is some $F$, or we can say, in other words, that $E$ is « a part » of « whole » $F$, i.e., $(E = E \cap F) = (E \subseteq F) = (E \cup F/E = F)$ (Jevons 1864). Several different partial identities are possible. If « a.e. » is « almost every », then $E(\mathbf{X},\mathbf{Y}) = F(\mathbf{X},\mathbf{Y})$ a.e.

$$= \begin{cases} F(\mathbf{X},\mathbf{Y}) & \text{if } \mathbf{R} < \mathbf{X}\,.\cap:\mathbf{P} < \mathbf{Y}\,. \\ f(\mathbf{X},\mathbf{Y}) & \text{if } \mathbf{R} > \mathbf{X}\,.\cup:\mathbf{P} > \mathbf{Y}\,. \end{cases} \Bigg|\, \infty > (\mathbf{R},\mathbf{P}) \in \underset{V=1}{\overset{T+I}{\times}} \mathbb{R}$$

$$= f(\mathbf{X},\mathbf{Y}) + [F(\mathbf{X},\mathbf{Y}) - f(\mathbf{X},\mathbf{Y})] \cdot \vartheta_{X-R\,|\,P-Y,\,1}\,,$$

which is to say, in other words, that $E = F$ « there where » $\mathbf{X}$ and $\mathbf{Y}$ are « sufficiently great ». This can be combined with « stability » $\widetilde{F}\mathbf{Z} = \{\,F\mathbf{Z}\,|\,F(\mathbf{Z} + d\mathbf{Z}) - F\mathbf{Z} - dF\mathbf{Z} \leq 0\,\}$; « stability » of $F\mathbf{Z}$ is $F\mathbf{Z} = \widetilde{F}\mathbf{Z} = \wp(F(\mathbf{Z}+d\mathbf{Z})\,;\,F\mathbf{Z},dF\mathbf{Z},J_M - J_N) \neq \varnothing$ (Waddington 1957).

Let $(E = f) \mapsto (\theta - C - P)(\mathbf{X},\mathbf{Y}) > 0 = \dot{\theta} = \dot{C} = \dot{P}$. We say of systems that even-rotation is steadiness of stable form-preserving change of change of state of systems; but steadiness and stability of their state is homeostasis and steadiness of stable form-preserving change of their state is homeorhesis. Once variable quantities are sufficient they are within a ball.

If $(E = F) \mapsto (\Theta - C - P)(\mathbf{X},\mathbf{Y}) \leq 0 = \dot{\Theta} = \dot{C} = \dot{P}$, then « attractive » center $C(\mathbf{X},\mathbf{Y})$ exists by way of « equilibrating tendency » $E = F$ a.e. that causes evolution of map $E$ into map $F$, and so « repulsive » center $c(\mathbf{X},\mathbf{Y}) \neq C(\mathbf{X},\mathbf{Y})$ exists, too.

All outputs, inputs, and parameters of systems can be thus perceived. Transformation $\rho$ of system $\varphi$ reversibly scales $\varphi \mathbf{X}$ where this is really necessary for existence of $(\mathbf{R}, \mathbf{P})$: « homeostasis » $F = \{ \widetilde{F} \mid E = \varphi \mid \dot{F} = 0 \}$ of systems is thus distinguishable from their « homeorhesis » $F = \{ \widetilde{F} \mid E = \partial \varphi / \partial T \mid \dot{F} = 0 \}$ and this is distinguishable itself from their « even-rotation in phase-space » $F = \{ \widetilde{F} \mid E = \partial^2 \varphi / (\partial T)^2 \mid \dot{F} = 0 \}$.

If so, then distinguishable special relations of variable quantities result in distinguishable equilibria of systems which are constructed from combinations of these variable quantities, e.g., $F$ is an « $E$-type equilibrium » if

$$\varphi \sum_{V=1}^{T} X_V \cdot \boldsymbol{\varepsilon}^V \underset{V}{<} \sum_{V=1}^{T} X_V \cdot \boldsymbol{\varepsilon}^V \ \bigg| \ X_V \in \mathbb{R} / \{0\} \mid X_W = X_v = 0 \ ,$$

but $F$ is a « $P$-type equilibrium » if

$$\varphi \sum_{V=1}^{T} X_V \cdot \boldsymbol{\varepsilon}^V \underset{V}{>} \sum_{V=1}^{T} X_V \cdot \boldsymbol{\varepsilon}^V \ \bigg| \ X_V \in \mathbb{R} / \{0\} \mid X_W = X_v = 0 \ ,$$

and so there where one or several observers prefer one equilibrium instead of another equilibrium what they really prefer there is one set of material relations instead of another set of material relations if systems whose equilibria they differently value are homologous. Absolute number of distinguishable $\mathbf{X}$s mapped within some ball to one state $\mathbf{Y}$ — in phase-space — is one different and more popular measure of its attraction; but economies are systems of actors whose actions consist of taking or leaving and so there each phase-space location $\mathbf{Y}$ is generated by infinitely many different quantitative configurations $\mathbf{X}$s (Fisher 1892; Cuhel 1907; Mises 1949). This truth limits possibility of so conventionally measuring attraction in economies or in dynamical systems that are homeomorphic to economies.

Equilibrium of economies thus results by a hypothetical process of attraction; but this steadiness is everywhere counterfactual and cannot actually exist (Hulsmann 2000).

Absence of uncertainty results from even-rotation of economies but this is not sufficient to cause even-rotation of economies. This remains true for every actual set of wants. Why <u>are</u> economies necessarily <u>actually</u> in state of disequilibrium and not such equilibrium? Let us replace that question — « When <u>are</u> economies <u>actually</u> in state of equilibrium? » — with a question that is equivalent to it: « When is ( **R** , **P** ) < ( **X** , **Y** ) ? » Action of actors is determined by their wants. When <u>are</u> these wants <u>actually</u> « sufficiently greatly » satisfied? Never. This possibly takes place there and only there where scarcity of goods is altogether in process of being diminished until it ultimately ceases to exist but this cannot actually happen: and why not? Preferences of actors are truly not limited over time, which is to say, in other words, that these actors always have wants that are not already satisfied.

If so, then aggregate over time of space-time events which satisfy wants that is necessary in order to satisfy all wants of actors is not finite itself and scarcity — difference of this set of events and set of events that actually take place in space-time — vanishes if and only if rarity of each event relative all other events vanishes and frequency of each event is possibly infinite. Energy conservation makes this physically impossible: and what cannot possibly take place also does not actually take place. Thus it remains fantastical.

Limitation by hypothesis of those preferences of actors which are not already satisfied (and their change over time) is solely what enables observers to postulate general equilibrium of economies but this general equilibrium is necessarily counterfactual for all possible preferences of actors: because these sets of wants are all really infinite and not limited. Observed divergence of actual behavior of actors from behavior of actors in evenly-rotating economies can neither measure distance of actions observed from those actions which result in state of greatest want-satisfaction — welfare — nor imply that this distance is greater than zero.

Thus observers cannot criticize an economic disequilibrium solely because it is an economic disequilibrium and not an economic equilibrium: because if they do so then they criticize what is possible for not being what is not possible and cannot actually exist, which is self-contradiction. Indeed they observers postulate an economic equilibrium counterfactually and only counterfactually in order to define and determine instantaneous profit and loss from perspective of preferences of consumers of products whose actions are limited by hypothesis to interval **B** of time whose instantaneous boundary $\partial$ **B** = [ $H$ , $K$ ] consists of present moment $H$ from their perspective and a sufficiently near future moment $K$ .

# References

**Bonnot de Condillac, Etienne. 1776.** *Commerce & Gouvernement*. Paris: Cellot, Jombert.

**Boskovich, Roger. 1755.** Supplementum. *Philosophiae Recentioris*. Rome: Palearini.

**Cantillon, Richard. 1755.** *Essai Sur la Nature du Commerce en General*. London: Gyles.

**Cuhel, Franz. 1907.** *Zur Lehre von den Bedurfnissen*. Innsbruck: Wagner.

**DeMorgan, Augustus. 1860.** *Syllabus of Proposed System of Logic.* London: Maberly, Walton.

**Destutt de Tracy, Antoine. 1817.** Treatise on Will & Its Effects. Translated by Jefferson, Thomas; Milligan, Joseph. *Treatise on Political Economy*. Georgetown: Milligan.

**Euler, Leonhard. 1741.** Letter of Leonhard Euler to Johann Bernoulli (in Hydraulica). *Opera Omnia, Volume 4*. Geneva: Bousquet.

**Fetter, Frank. April 1910.** Phenomena of Economic Dynamics. *American Economic Association Quarterly* 11(1):130-135.

**Fisher, Irving. April 1892.** Mathematical Investigations in Theory of Value & Prices. *Transactions of Connecticut Academy of Arts & Sciences* 9(1):1-124.

**Gossen, Hermann. 1854.** *Entwicklung der Gesetze des Menschlichen Verkehrs*. Braunschweig: Vieweg.

**Hayek, Friedrich von. 1937.** Economics & Knowledge. *Economica* 4(1):33-54.

_____. **September 1945.** Use of Knowledge in Society. *American Economic Review* 35(4): 519-530.

**Hulsmann, Guido. Winter 2000.** Equilibrium Analysis. *Quarterly Journal of Austrian Economics* 3(4):3-51.

**Hutt, William. 1939.** *Theory of Idle Resources*. London: Cape.

_____. **1956.** Yield From Money Held. *Freedom & Free Enterprise.* Princeton: Van Nostrand.

**Hutton, James. 1794a.** *Investigation of Principles of Knowledge, Volume 1*. Edinburgh: Cadell, Strahan.

_____. **1794b.** *Investigation of Principles of Knowledge, Volume 2*. Edinburgh: Cadell, Strahan.

**Knight, Frank. 1921.** *Risk, Uncertainty, Profit*. Boston: Houghton-Mifflin.

**Kothe, Gottfried. December 1952.** Randverteilungen Analytischer Funktionen. *Mathematische Zeitschrift* 57(1):13-33.

**Leibniz, Gottfried von. [November 1676] 2001.** Pacidius Philalethi. *Labyrinth of the Continuum*. New Haven: Yale University Press.

_____. **April 1695.** Specimen Dynamicum. *Acta Eruditorum* 14(4):145-157.

_____. **1717.** *Papers which Passed between Leibniz & Clarke relating to Principles of Natural Philosophy & Religion*. London: Knapton.

**Longfield, Mountifort. 1834.** *Lectures on Political Economy*. Dublin: Milliken.

**Lotka, Alfred. 1925.** *Elements of Physical Biology*. Baltimore: Wilkins; Williams.

**MacLane, Saunders. 1971.** *Categories for Working Mathematician*. New York: Springer.

**Mayer, Julius. May 1842.** Bemerkungen uber die Krafte der Unbelebten Natur. *Annalen der Chemie und Pharmacie* 42(2):233-240.

**Menger, Carl. 1871.** *Grundsatze de Volkswirthschaftslehre*. Vienna: Braumuller.

**Menger, Karl. 1944.** *Algebra of Analysis*. Notre Dame: University of Notre Dame Press.

⎯⎯⎯. **1952.** *Calculus*. Chicago: Illinois Institute of Technology Bookstore.

⎯⎯⎯. **1953.** *Calculus* (Revised Edition). Chicago: Illinois Institute of Technology Bookstore.

**Mises, Ludwig von. 1912.** *Theorie des Geldes und der Umlaufsmittel*. Munich: Duncker, Humblot.

⎯⎯⎯. **1949.** *Human Action*. New Haven: Yale University Press.

⎯⎯⎯. **1952.** Profit & Loss. *Planning for Freedom*. South Holland: Libertarian Press.

⎯⎯⎯. **1957.** *Theory & History*. New Haven: Yale University Press.

**North, Dudley. 1691.** *Discourses Upon Trade*. London: Basset.

**Rosen, Robert. March 1968.** Physico-Chemical Description of Biological Activity. *Journal of Theoretical* Biology 18(3):380-386.

⎯⎯⎯. **1985.** *Anticipatory Systems*. Oxford: Pergamon.

⎯⎯⎯. **1991.** *Life Itself*. New York: Columbia University Press.

**Rothbard, Murray. 1979.** Myth of Efficiency. *Time, Uncertainty, Dis-Equilibrium*. Lexington: Heath.

**Sato, Mikio. March 1958.** Generalization of Concept of Functions, Part 1. *Proceedings of Japan Academy* 34(3):126-130.

⎯⎯⎯. **March 1959.** Theory of Hyperfunctions, Part 1. *Journal of Faculty of Science of University of Tokyo* 8(1):139-193.

⎯⎯⎯. **March 1960.** Theory of Hyperfunctions, Part 2. *Journal of Faculty of Science of University of Tokyo* 8(2):387-437.

**Spencer, Herbert. 1864.** *Principles of Biology, Volume 1*. London: Norgate; Williams.

**Tait, Peter; Thomson, William. 1867.** *Treatise on Natural Philosophy*. Oxford: Oxford University Press.

**Thom, Rene. [1981] 1983.** *Mathematical Models of Morphogenesis*. Translated by Brookes, William; Rand, David. Chichestor: Horwood.

**Waddington, Conrad. 1940.** *Organizers & Genes*. London: Allen, Unwin.

**Waddington, Conrad. 1957.** *Strategy of Genes*. London: Allen, Unwin.

**Wicksteed, Philip. 1910.** *Common Sense of Political Economy*. London: Macmillan.